\documentclass[aps,prd,10pt,twocolumn,superscriptaddress,showpacs,floatfix]{revtex4-1}
\usepackage{textcomp,gensymb}
\usepackage{hyperref}
\usepackage{graphicx}
\usepackage{amsfonts,amsmath,amssymb,bm,bbm}
\usepackage{color,soul}
\usepackage{slashed}
\usepackage[toc,page]{appendix}
\usepackage{flushend}
\usepackage{physics}
\usepackage{graphicx}
\usepackage{dcolumn}
\usepackage{float}
\usepackage{fontawesome} % used for the github icon

\def\apj{ApJ}%
\def\apjl{ApJ}%
\def\jcap{JCAP}%
\def\prd{Phys.\ Rev.\ D}%

\graphicspath{{./figures/}}

\hypersetup{
    pdfnewwindow=true,      
    colorlinks=true,       
    linkcolor=blue,          
    citecolor=blue,        
    filecolor=blue,      
    urlcolor=blue        
}

\newcommand{\sv}{\langle \sigma v \rangle}

\begin{document}

\title{The Morphology of Exciting Dark Matter and the Galactic 511 keV Signal}

\author{Christopher V. Cappiello}
\email{cvc1@queensu.ca}
\affiliation{Department of Physics, Engineering Physics, and Astronomy, Queen's University, Kingston, ON, K7L 3N6, Canada}
\affiliation{Arthur B. McDonald Canadian Astroparticle Physics Research Institute, Kingston, ON, K7L 3N6, Canada}
\affiliation{Perimeter Institute for Theoretical Physics, Waterloo, ON, N2L 2Y5, Canada}

\author{Michael Jafs}
\email{michael.jafs@mail.mcgill.ca}
\affiliation{Department of Physics, Engineering Physics, and Astronomy, Queen's University, Kingston, ON, K7L 3N6, Canada}
\affiliation{Department of Physics, McGill University, 3600 Rue University, Montr\'{e}al, Qu\'{e}bec, Canada, H3A 2T8}

\author{Aaron C. Vincent}
\email{aaron.vincent@queensu.ca}
\affiliation{Department of Physics, Engineering Physics, and Astronomy, Queen's University, Kingston, ON, K7L 3N6, Canada}
\affiliation{Arthur B. McDonald Canadian Astroparticle Physics Research Institute, Kingston, ON, K7L 3N6, Canada}
\affiliation{Perimeter Institute for Theoretical Physics, Waterloo, ON, N2L 2Y5, Canada}

\date{\today}

\begin{abstract}
We study the morphology of the 511 keV signal that could be produced by exciting dark matter (XDM) in the Milky Way. In this model, collisions between dark matter particles excite the dark matter to a state that can then decay back to the ground state, releasing an electron-positron pair. These electrons and positrons would then annihilate, producing 511 keV photons that could explain the 511 keV signal seen by INTEGRAL at the Galactic Center. We compare the resulting flux with the most recent INTEGRAL data, performing the first full statistical analysis of the exciting dark matter model. We focus on exciting dark matter in the mass and cross section ranges 100 GeV $\lesssim m_{\chi} \lesssim$ 3 TeV and $10^{-19}$ cm$^3$ s$^{-1} \lesssim \langle \sigma v \rangle \lesssim 10^{-16}$ cm$^3$ s$^{-1}$. We show that exciting dark matter can provide a significantly better fit than the simpler case of annihilating dark matter, with $\Delta\chi^2 > 16$ for all but one of the density profiles we consider. 
\end{abstract}

\maketitle

%%%%%%%%%%%%%%%%%%%%%%%%%%%%%%%%%%%%%%%%%%%%%%%%%%%%%%%%%%%%%%%%%%
%%%%%%%%%%%%%%%%%%%%%%%%%%%%%%%%%%%%%%%%%%%%%%%%%%%%%%%%%%%%%%%%%%

\section{Introduction}

More than 50 years after it was first observed~\cite{1972ApJ...172L...1J}, the 511 keV gamma-ray line at the Galactic Center remains unexplained~\cite{Prantzos:2010wi,Leane:2022bfm,Siegert:2023wus}. Despite the difficulties in characterizing a flux of $\sim$ MeV gamma rays, successive analyses of data from CGRO/OSSE \cite{1997ApJ...491..725P} and then INTEGRAL/SPI \cite{Bouchet_2010,WEIDENSPOINTNER2008454,2011ApJ...739...29B,Siegert:2019tus} have led to a consistent picture: a significant 511 keV signal from the galactic disk, combined with a large, spherically symmetric ``bulge'' signal centered on the the galactic center and extending roughly 10 degrees above and below the galactic plane. The bulge and disk both contribute about 10$^{-3}$ ph cm$^{-2}$ s$^{-1}$, and the spectra are consistent with a positronium formation rate around unity, pointing at low-energy ($\lesssim 10$ MeV) injection of positrons into the interstellar medium (ISM). Much of the disk emission can be attributed to $\beta^+$ decays of $^{26}$Al, $^{44}$Ti and $^{56}$Ni, and indeed, the concurrent 1809 keV gamma signal from $^{26}$Al $\rightarrow$ $^{26}$Mg$^*$ + $\beta^+$, $^{26}$Mg$^*$ $\rightarrow$ $^{26}$Mg $+ \gamma$ provides an additional handle on radioactive isotope contribution. However, the bulge contribution is uncorrelated with the morphology of observations in any other band \cite{Prantzos:2010wi}, though old stellar populations \cite{Crocker:2016zzt} or even extragalactic positrons \cite{Vincent:2015boa} have been suggested as possible sources. Nonetheless, the lack of a confirmed counterpart in other bands and the presence of this significant spheroidal component have motivated numerous attempts to identify the excess as a signal of dark matter~\cite{Boehm:2003bt,Boehm:2003ha,Hooper:2004qf,Picciotto:2004rp,Gunion:2005rw,Chun:2006ss,Finkbeiner:2007kk,Huh:2007zw,Pospelov:2007mp,Pospelov:2007xh,Arkani-Hamed:2008hhe,Chen:2009av,Chen:2009dm,Finkbeiner:2009mi,Cline:2010kv,Cline:2012yx,Vincent:2012an,Farzan:2017hol,Jia:2017iyc,Farzan:2020llg,Ema:2020fit} or other physics beyond the Standard Model, such as cosmic strings~\cite{Ferrer:2005xva,Ferrer:2006pf}. 

To successfully explain the galactic center signal, a model must correctly predict: 1) the total flux, 2) the line spectrum, and 3) the observed morphology. Annihilating dark matter of mass $m_\chi$ requires an annihilation cross section $\sv/m_\chi^2 \sim 10^{-25}$ cm$^3$ s$^{-1}$ GeV$^{-2}$ \cite{Vincent:2012an}. Obtaining the correct spectrum requires $m_\chi \lesssim 10$ MeV, otherwise final-state radiation and in-flight annihilation risk overproducing photons above 511 keV \cite{Ascasibar:2005rw,Beacom:2005qv,Sizun:2007ds}. Combining these constraints significantly restricts the parameter space, leading to cross sections $\sv \sim 10^{-31}$, which would either severely overproduce dark matter in the early Universe, or require additional annihilation to light products, in conflict with big bang nucleosynthesis (BBN) or recombination observations \cite{Wilkinson:2016gsy}. Finally, the morphology of the signal predicted from dark matter annihilation can end up being \textit{too} cuspy, based on the more common models of the Milky Way halo \cite{Vincent:2012an}. 

A class of dark matter models that addresses these drawbacks  is exciting dark matter (XDM), in which collisions between dark matter particles excite the dark matter into a state which can then decay to a lower-energy state via emission of an electron-positron pair~\cite{Finkbeiner:2007kk}. Such a model has the advantage of producing relatively low-energy electron-positron pairs even if the dark matter mass is in the GeV--TeV range characteristic of a WIMP. The velocity threshold required to produce an excitation also suppresses $e^+$ production very close to the galactic centre where the velocity dispersion is lower, leading to a less cuspy predicted signal with respect to DM annihilation. Much work has focused on whether such a model could produce the correct 511-keV flux~\cite{Pospelov:2007xh,Arkani-Hamed:2008hhe,Chen:2009av,Finkbeiner:2009mi,Cline:2012yx}, and while some analyses of the morphology were included, these were performed before state-of-the art kinematic analyses of the Milky Way dynamics were available, did not use data based on the latest SPI measurements, and most did not attempt to quantify a goodness of fit between model predictions and available data.

In this work, we thus compare the morphology of the predicted XDM signal with the signal produced by more traditional annihilating dark matter. By testing these predictions against the most recent INTEGRAL data, we show that exciting dark matter provides a significantly better fit, for most realistic choices of the dark matter density profile.

\section{Dark Matter Models}

\subsection{Exciting Dark Matter}

In the XDM scenario, the DM density is primarily composed of a stable state $\chi_0$, which can be excited to a higher-energy state $\chi_1$ and subsequently decay, producing an electron-positron pair. The excitation is induced by a collision between two DM particles, in contrast with the case of inelastic DM, which usually refers to collisions between DM and Standard Model particles in a detector (see e.g.~\cite{Tucker-Smith:2001myb}). To produce an electron-positron pair, the mass splitting must be at least twice the electron mass:
\begin{equation}
    \delta m \equiv m_1 - m_0 \geq 2m_e\,.
\end{equation}
In this work, we assume that $\delta m = 2m_e$.

This excitation is only possible if the center-of-mass energy is at least $2m_0 + \delta m$, meaning that the relative velocity between the two DM particles must be above a threshold:
\begin{equation}
    v_{th} = \sqrt{4\delta m/m_0}\,.
\end{equation}
This threshold introduces a velocity dependence to the cross section for excitation to occur~\cite{Finkbeiner:2007kk}:
\begin{equation}
\sigma v_{rel}=
    \begin{cases}
        \sigma_{mr}\sqrt{v_{rel}^2 - v_{th}^2} & v_{rel}>v_{th}\\
        0 & v_{rel}\leq v_{th}\,,
    \end{cases}
\end{equation}
where $\sigma_{mr}$ is the cross section in the moderately-relativistic limit. As we will see below, the dependence on relative velocity translates into a dependence on galactocentric radius, suppressing the excitation rate near the Galactic Center.

\subsection{Annihilating Dark Matter}

Simpler than the above scenario is the case where a DM pair annihilates to an electron-positron pair. In this case, there is no threshold velocity, and no need for multiple DM states with a specific mass splitting. However, if the DM is very heavy, then an additional gamma-ray component above 511 keV would be expected from in-flight annihilation and final-state radiation, and the absence of such a signal limits the annihilating DM mass to $m \lesssim$ 3-10 MeV~\cite{Beacom:2005qv}, which puts this scenario in tension with cosmological bounds on the minimum mass of thermal DM~\cite{Boehm:2013jpa,Wilkinson:2016gsy,Depta:2019lbe,Sabti:2019mhn,An:2022sva,Chu:2022xuh} (although see Ref.~\cite{Berlin:2017ftj}). We take annihilating DM as an alternative to XDM, and compare how well the morphology of both signals fits observations.

\section{Signal Rate Calculation}

\subsection{Dark Matter Signal}

Assuming that the electron-positron production and annihilation are in equilibrium, that each collision yields one $e^+e^-$ pair, and that the electrons and positrons do not travel far before annihilating, the flux per steradian of 511 keV gamma rays for a given galactic longitude $l$ and latitude $b$ is given by the line of sight integral  ~\cite{Finkbeiner:2007kk} 
\begin{equation}\label{eq:511flux}
    \Phi_{511}(b,l) = \frac{1 - 0.75f_p}{4\pi}\int_0^{\infty}\textrm{d}x\langle\sigma v\rangle \left(\frac{\rho(r)}{m_{\chi}}\right)^2\,,
\end{equation}
where $f_p = 0.967$ is the positronium fraction. Because of the form of Eq.~\eqref{eq:bigsv}, and the radial dependence of the dark matter halo dispersion velocity, the thermally-averaged cross section $\langle\sigma v\rangle $ is a function of the galactocentric radius $r(x,l,b)$.

We consider two types of profile for the DM density distribution: a generalized Navarro-Frenk-White (gNFW) profile~\cite{Zhao:1995cp,1997ApJ...490..493N} and an Einasto profile~\cite{1965TrAlm...5...87E}, given respectively by
\begin{equation}
    \rho_{NFW}(r) = \frac{2^{3-\gamma}\rho_s}{(r/r_s)^{\gamma}(1+r/r_s)^{3-\gamma}}\,,
\end{equation}
\begin{equation}
    \rho_{Ein}(r) = \rho_s\, \textrm{exp}\left[-\left(\frac{2}{\alpha}\left(\frac{r}{r_s}\right)^{\alpha} - 1\right)\right]\,.
\end{equation}
For both profiles, we use the best-fit values of the local density, slope, and scale radius obtained by Ref.~\cite{2019JCAP...10..037D} from rotation curve data of the Milky Way. We note that these data only extend down to radii of about 5 kpc, making it difficult to constrain the DM density at the Galactic Center, where the 511-keV signal that we compute should be strongest. For this reason, we consider four different models of the DM density profile: one Einasto and one gNFW profile for each of the two baryonic models used in~Ref~\cite{2019JCAP...10..037D}, denoted B1 and B2. The values of these parameters are shown in Table~\ref{tab:profiles}.

\begin{table}
\begin{tabular}{ c c c c } 
 \hline
Model & $\rho_{DM}$ [GeV/cm$^3$] & $r_s$ [kpc] & $\alpha,\gamma$ \\ [0.5ex] 
 \hline\hline
 gNFW1 & 0.30 & 9 & 1.2 \\ 
 Einasto1 & 0.30 & 11 & 0.11 \\
 gNFW2 & 0.39 & 8.1 & 1.3 \\ 
 Einasto2 & 0.38 & 9.2 & 0.18 \\
 \hline \hline
\end{tabular}
\caption{\label{tab:profiles}List of DM density profile models, with parameters from the B1 and B2 models of Ref.~\cite{2019JCAP...10..037D}.}
\end{table}

To compute the radial dependence of the thermally averaged cross section, we require the relative velocity distribution of DM as a function of its position in the galaxy. We model the velocity distribution at any point in the galaxy as a Maxwellian, as in the Standard Halo Model~\cite{Drukier:1986tm}:
\begin{equation}
    f(\vec{v}) = \frac{1}{N_{esc}(2\pi\sigma_v^2)^{3/2}}\,e^{-\frac{\vec{v}^2}{2\sigma_v^2}}\,\Theta(v_{esc} - v)\,,
\end{equation}
where $N_{esc}$ is a normalization constant accounting for the cutoff at $v_{esc}$. Neglecting the escape velocity cutoff, the distribution of relative velocities between 2 particles takes the same form, with the replacement $\sigma_{rel} = \sqrt{2}\sigma_v$ (see, e.g., Ref.~\cite{2009PhRvD..79h3525R}). When we include the cutoff, as well as the velocity dependence of the cross section, we get a more complicated form for the thermally averaged cross section:

\begin{widetext}
    \begin{align}
    \langle\sigma v\rangle &= \langle\sigma v\rangle_{mr} 
    2\pi\left(\frac{1}{N_{esc}}\frac{1}{(2\pi\sigma_v^2)^{3/2}}\right)^2 \nonumber \\ & \times \int_{v_{th}}^{2v_{esc}}\textrm{d}v_{rel}\sqrt{v_{rel}^2 - v_{th}^2}e^{-\frac{2v_{esc}^2+v_{rel}^2}{2\sigma_v^2}}\sigma_v^3  
    \left(2\sigma_v(e^{\frac{v_{rel}^2}{2\sigma_v^2}} - e^{\frac{v_{esc}v_{rel}}{2\sigma_v^2}}) + \sqrt{\pi}v_{rel}e^{\frac{4v_{esc}^2 + v_{rel}^2}{4\sigma_v^2}}\textrm{Erf}\left(\frac{2v_{esc}-v_{rel}}{2\sigma_v}\right)\right)\,. \label{eq:bigsv}
    \end{align}
\end{widetext}

In order to use Eq. \eqref{eq:bigsv}, we need $\sigma_v$ and $v_{esc}$ as a function of position in the galaxy. Assuming spherical symmetry, $\sigma_v$ can be computed using the Jeans Equation~\cite{2009PhRvD..79h3525R}:
\begin{equation}
    \sigma_v^2(r) = \frac{1}{\rho(r)}\int_{\infty}^r\rho(r)\frac{\textrm{d}\phi}{dr}dr\,,
    \label{eq:jeans}
\end{equation}
where $\rho(r)$ is the DM density and $\phi(r)$ is the gravitational potential of the galaxy (DM and baryonic components). The gravitational potential due to the DM is just 
\begin{equation}
    \phi_{DM}(r) = \int_{\infty}^r\frac{GM_{enc}}{r'^2}dr' = 4\pi G\int_{\infty}^r\frac{\textrm{d}r'}{r'^2}\int_0^{r'}\textrm{d}r''\rho(r'')r''^2\,.
\end{equation}
 To make use of Eq.~\eqref{eq:jeans}, we employ spherically symmetrized potentials for the baryonic components, as in Refs.~\cite{Strigari:2009zb,Pato:2012fw,Boddy:2018ike}. 
We include the bulge and disk potentials used in Ref.~\cite{Boddy:2018ike},
\begin{equation}
    \phi_{Bulge}(r) = -\frac{GM_b}{r+c_0}\,,
\end{equation}
\begin{equation}
    \phi_{Disk}(r) = -\frac{GM_d}{r}(1 - e^{-r/b_d})\,,
\end{equation}
where $M_b = 1.5 \times 10^{10}$ M$_{\odot}$ is the mass of the bulge, $c_0 = 0.6$ kpc is the bulge scale radius, $M_d = 7 \times 10^{10}$ M$_{\odot}$ is the mass of the disk, and $b_d = 4$ kpc is the disk scale radius. Our final model for the gravitational potential is then
\begin{equation}
    \phi(r) = \phi_{DM}(r) + \phi_{Bulge}(r) + \phi_{Disk}(r)\,.
\end{equation}
Once we have the potential, the escape velocity is also easy to compute~\cite{Boddy:2018ike}:
\begin{equation}
    v_{esc}(r) = \sqrt{-2\phi(r)}\,.
\end{equation}
With all these pieces, we can now use Eq.~\eqref{eq:511flux} to compute the 511 keV flux due to both exciting and annihilating DM.

\subsection{Disk Flux}

On top of a possible 511 keV signal from DM, we know that there should exist 511 keV emission from the galactic disk, due to radioactive isotopes such as $^{26}$Al. We model the disk emission using the Robin young disk model~\cite{Robin:2003uus}:
\begin{equation}
    \dot{n}(r) = \dot{n}_0\left(e^{-\frac{a^2}{R_0^2}} - e^{-\frac{a^2}{R_i^2}}\right)\,,
\end{equation}
where $R_0$ = 5 kpc, $R_i$ = 3 kpc, and 
\begin{equation}
    a = \sqrt{x^2 + y^2 + z^2/\epsilon^2}\,.
\end{equation}
Here $\epsilon$ parametrizes the disk scale height, and in our fit is allowed to vary between 0.01 and 0.03.

$^{26}$Al $\beta^+$ decay leads to an excited state of $^{26}$Mg, which yields a deexcitation gamma line at 1809 keV. Based on 1809 keV measurements, the corresponding total 511 keV emission due to $^{26}$Al is (7.33 $\pm$ 0.89) $\times 10^{-4}$ ph cm$^{-2}$ s$^{-1}$~\cite{Diehl:2005py,Vincent:2012an}. The disk emission could be larger than this, due to other elements (which do not come with such a convenient tracer line), but it should not be smaller. We use this value as an additional constraint on the value of $\dot{n}_0$, as described below.

\section{Data Fitting}

We compare the computed 511 keV flux from the disk and DM to the data from SPI aboard the INTEGRAL satellite used in Refs.~\cite{Siegert:2015knp,Siegert:2019tus} (data provided by the authors). The data are based on 11 years of exposure, with some gaps due to solar activity, calibration or annealing. Recent analyses have made extensive use of the $\sim$3-year data presented in Ref.~\cite{Bouchet_2010}. We stress that more exposure is particularly important in teasing out the morphology of such a signal because of the shallow intensity gradients involved and the large isotropic (instrumental) background 511 keV event rate.

\begin{figure}
    \centering
    \includegraphics[width=\columnwidth]{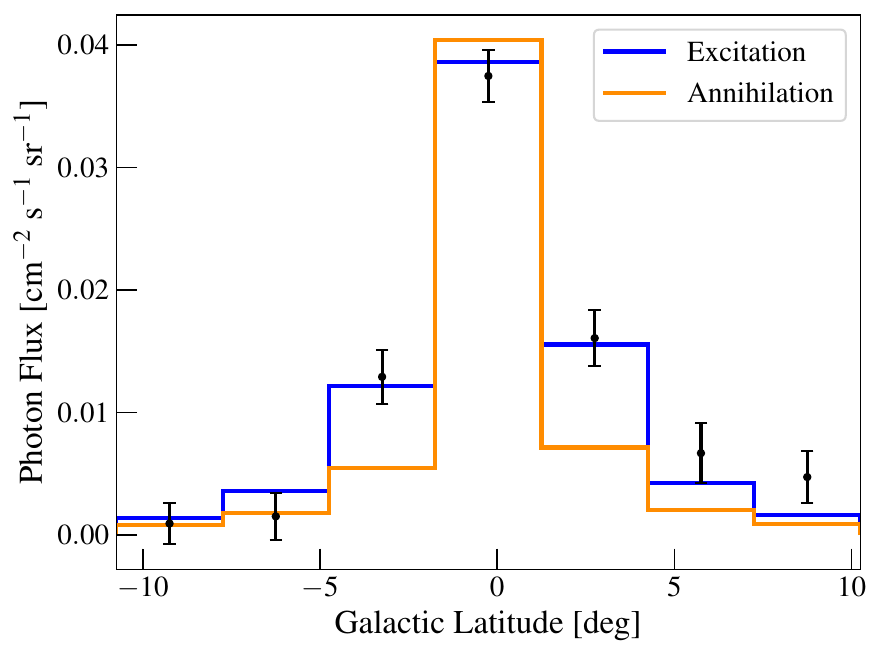}
    \caption{The central latitudinal profile of 511-keV emission, with data provided by the authors of Ref.~\cite{Siegert:2015knp}. Blue and orange are the best fits using the gNFW2 profile, for exciting and annihilating DM, respectively. In both plots $m_{\chi}$ = 481 GeV. The fits appear asymmetric only because the data are not centered quite on the Galactic Center, but rather at -0.25 degrees~\cite{Siegert:2019tus}.}
    \label{fig:latprofile}
\end{figure}

\begin{figure}
    \centering
    \includegraphics[width=\columnwidth]{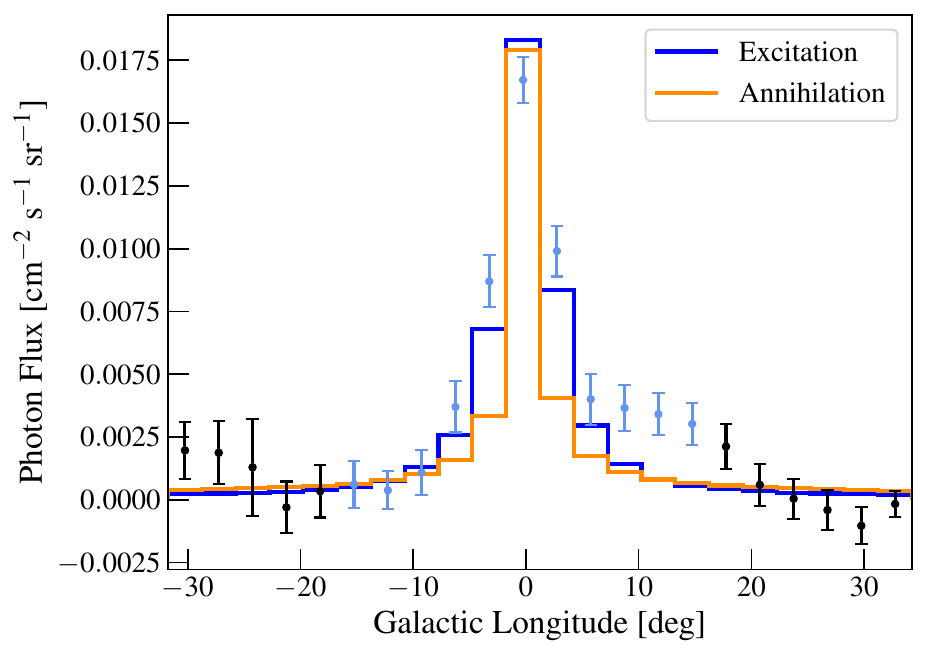}
    \caption{Same as Fig.~\ref{fig:latprofile}, but showing the longitudinal profile. The lower normalization in the central bin compared to Fig.~\ref{fig:latprofile} is due to the longitudinal bins covering a larger solid angle, and thus the central bin being less dominated by the peak at the Galactic Center. As above, the slightly asymmetric shape is due only to the data being centered at -0.25 degrees.}
    \label{fig:lonprofile}
\end{figure}

Here we describe the dataset used in terms of galactic longitude $l$ and latitude $b$.
At longitudes near the galactic center, we consider three latitudinal profiles, for the longitude ranges $-15.25^{\circ} < l < -6.25^{\circ}$, $-5.25^{\circ} < l < 3.75^{\circ}$, and $4.75^{\circ} < l < 13.75^{\circ}$. Each profile consists of 7, 3-degree bins in the range from $-10.75^{\circ} < b < 10.25^{\circ}$. Data points for the central profile are shown in Fig. \ref{fig:latprofile}.  For longitudes further from the galactic center, we no longer have latitude profiles. Instead, we have a longitudinal profile integrated over all latitudes $-10.75^{\circ} < b < 10.25^{\circ}$, in 3-degree longitude bins covering the range $-31.75^{\circ} < l < 34.25^{\circ}$. To avoid overlap with the latitudinal profiles, we use the longitudinal profile only in the ranges $-31.75^{\circ} < l < -16.75^{\circ}$ and $16.25^{\circ} < l < 34.25^{\circ}$.  This is shown in Fig. \ref{fig:lonprofile}, where we have coloured the data points that do not contribute to the likelihood analysis in periwinkle blue.

For each of the DM profiles described in Table~\ref{tab:profiles}, we fit the DM plus disk flux to the data described above. We define a modified $\chi^2$, which penalizes the fit if it requires a disk flux smaller than the disk flux inferred from 1809 keV data:

\begin{equation}
    \chi^2 = \sum_i \frac{(y_{i,fit} - y_i)^2}{\sigma_i^2} + \frac{(D_{fit} - D_{Al})^2}{\sigma_{Al}^2}\Theta(D_{Al} - D_{fit})\,,
\end{equation}
where $D_{fit}$ is the fit value of the all-sky disk flux, $D_{Al} = 7.33 \times 10^{-4}$ ph cm$^{-2}$ s$^{-1}$ is the inferred flux from 1809 keV data and $\sigma_{Al} = 8.9 \times 10^{-5}$ ph cm$^{-2}$ s$^{-1}$ is the uncertainty on the aluminum contribution. The step function ensures that this term only penalizes the fit for a disk contribution that is too small, as it could be larger due to contributions from other $\beta^+$-decaying isotopes. $y_i$ and $y_{i,fit}$ are the observed and fit values of the flux in a given bin, and $\sigma_i$ the uncertainty on the flux in that bin. 

We find the minimum $\chi^2$ by varying the disk height parameter within the range $0.01 \leq \epsilon \leq 0.03$, and also varying the normalization of the disk flux. Below, we present our results in the plane of $m_{\chi}$ and $\langle\sigma_{mr}v\rangle$. 

\section{Results}

Table~\ref{tab:fits} shows the best fit $\chi^2$, and the corresponding values of $m_{\chi}$ and $\langle\sigma_{mr}v\rangle$, for the four DM density profiles we consider, each for both annihilating DM and XDM. In order to properly compare the XDM and annihilating DM models, we note that the XDM model has one additional degree of freedom, because in the annihilating case the mass and cross section are degenerate, while for XDM they are not. Assuming Wilks' theorem holds (the two models are nested, in that the high-mass limit of the XDM model maps to the annihilation model with a single degree of freedom), then the $\Delta\chi^2$ represents twice the log-likelihood ratio and should follow a $\chi^2$ distribution with one degree of freedom. From this we find, for example, that $\Delta\chi^2 = 16$ corresponds to a 4$\sigma$ difference between the two models. For three of the four profiles (both gNFW profiles and one Einasto profile), XDM provides a better fit than annihilating DM by $\Delta\chi^2 > 16$, thus corresponding to a greater-than $4\sigma$ improvement. The difference for the gNFW profiles is not surprising given the results of Ref.~\cite{Vincent:2012an}, which found that annihilating DM with a gNFW profile could not provide a good fit to the observed longitudinal profile. Ref.~\cite{Vincent:2012an} also found that an Einasto profile (assuming again annihilating DM) could fit the observed excess much better than a gNFW profile could. While the Einasto profile of model B2 fits the data nearly as well with annihilating DM as with XDM, this is not true for the cuspier Einasto profile of model B1. For the B1 Einasto profile, the annihilating DM case is actually a worse fit than the annihilating B1 gNFW profile, and is significantly worse than any profile with XDM.

Figures~\ref{fig:latprofile} and~\ref{fig:lonprofile} illustrate the difference between annihilation and XDM fits for the B2 gNFW model from Ref.~\cite{2019JCAP...10..037D}. We find that the case of annihilating DM is too cuspy to fit the data well, with a sharp peak at zero longitude. However, because for XDM the rate of electron-positron pair production is kinematically suppressed in the Galactic Center, the profile is less sharply peaked, and fits the observed data noticeably better.

\begin{figure}
    \centering
    \includegraphics[width=\columnwidth]{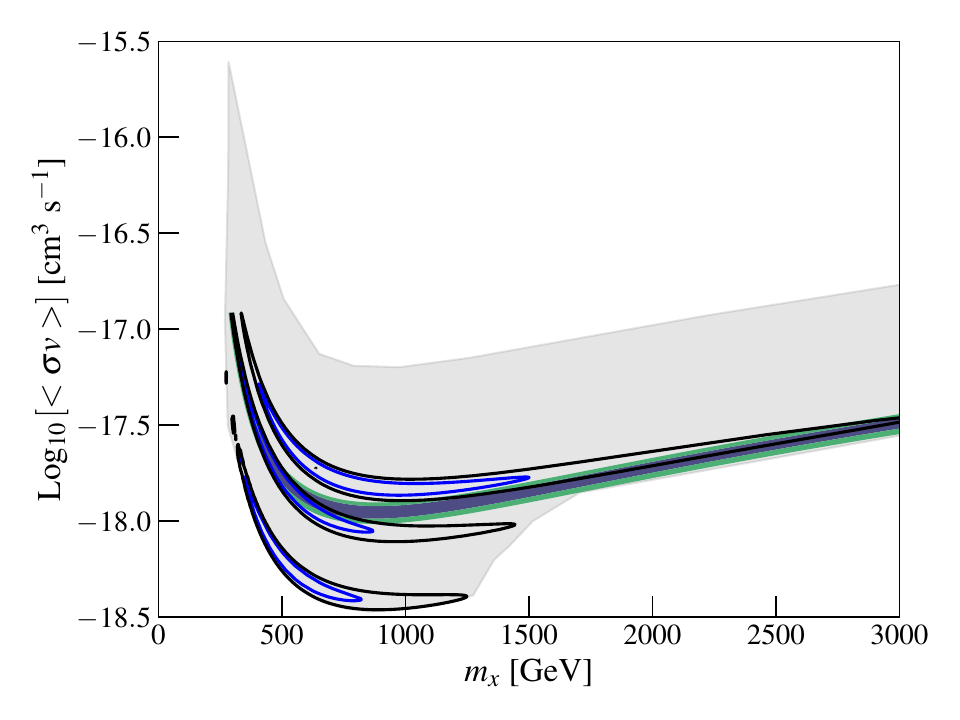}
    \caption{1 and 2$\sigma$ contours for XDM properties with the four density profiles we consider. For the Einasto2 model, which produces the best overall fit, we show filled contours. For the other models, we show unfilled curves. The gray region bounds the best fit region for these plus 12 other density profiles, showing the range of XDM models that could yield the observed 511 keV signal (see text).}
    \label{fig:sigmas}
\end{figure}

In Fig.~\ref{fig:sigmas}, we show contours corresponding to 1- and 2-$\sigma$ best fit regions for the four different density profiles we consider. The filled (dark green) contours are for the Einasto2 model, the one which gives the overall lowest $\chi^2$. To estimate how the uncertainties in profile parameters (namely slope and scale radius) affect the position of these best fit contours, we also considered 12 other NFW profiles with parameters drawn from within the 2-$\sigma$ contour shown in Ref.~\cite{2019JCAP...10..037D} for their B2 baryonic model. We included only profiles that fit the 511-keV data with $\chi^2 < 70$. The gray region in Fig.~\ref{fig:sigmas} encompasses all of the corresponding 2-$\sigma$ contours.

\begin{table}
\begin{tabular}{ c c c c } 
 \hline
Model & $\chi^2$ & $m_{\chi}$ [GeV] & $\langle\sigma_{mr}v\rangle$ [cm$^3$ s$^{-1}$] \\ [0.5ex] 
 \hline\hline 
 gNFW1, Ann & 79.9 & -- & $1.6 \times 10^{-25}\, (m_{\chi}/\textrm{GeV})^2$ \\
 gNFW1, XDM & 62.4 & 637 & $1.9 \times 10^{-18}$ \\ 
 \hline
 Einasto1, Ann & 87.7 & -- & $9.7 \times 10^{-26}\, (m_{\chi}/\textrm{GeV})^2$ \\
 Einasto1, XDM & 65.6 & 486 & $1.8 \times 10^{-18}$ \\
 \hline
 gNFW2, Ann & 99.8 & -- & $3.8 \times 10^{-26}\, (m_{\chi}/\textrm{GeV})^2$ \\
 gNFW2, XDM & 63.7 & 481 & $7.1 \times 10^{-19}$ \\
 \hline
 Einasto2, Ann & 62.3 & -- & $2.0 \times 10^{-25}\, (m_{\chi}/\textrm{GeV})^2$ \\
 Einasto2, XDM & 61.2 & 547 & $1.5 \times 10^{-18}$ \\
 \hline \hline
\end{tabular}
\caption{\label{tab:fits}$\chi^2$ and best fit DM properties for the different DM profiles considered, for annihilating and exciting DM. For the annihilation case, $m_{\chi}$ and $\langle\sigma_{mr}v\rangle$ are degenerate, so we give the best fit value of $\langle\sigma_{mr}v\rangle$ in terms of $m_{\chi}$ in the last column.}
\end{table}

\section{Conclusions}
We have studied the morphology of the 511-keV signal observed by SPI aboard the Integral satellite, performing the first statistical analysis to compare the fits of annihilating and exciting dark matter to this data. Previous work, which assumed annihilating dark matter, found that a generalized NFW profile is far too cuspy to fit the observed 511-keV data, and that an Einasto profile fit the data much better. On the other hand, we find that the ability of annihilating dark matter to fit the excess depends strongly on the parameters of the assumed density profile, and that an Einasto profile does not always give a better fit. Furthermore, we find that the fit for exciting dark matter is much less sensitive to the choice of profile than for annihilating dark matter, and in many cases is significantly better than the fit with annihilating dark matter. If one assumes that the Milky Way dark matter halo follows a generalized NFW profile, our results thus improve the viability of a dark matter explanation for the 511-keV excess. Our results also generally favor exciting dark matter over annihilating dark matter, which was already in tension with cosmological constraints on light particles.

As mentioned above, several other dark matter models have been put forward to explain the 511-keV excess. A statistical analysis of all these alternatives is beyond the scope of this work, but we note here some qualitative differences between our model and others (for a review of the morphology of several Standard Model explanations, see Ref.~\cite{Prantzos:2010wi}). The model we consider is quite similar to the endothermic models explored in Refs.~\cite{Chen:2009dm,Finkbeiner:2009mi,Cline:2010kv}, the most striking difference being that they require a lower threshold velocity for a given mass, which should lead them to predict a somewhat cuspier emission profile than we do. Refs.~\cite{Cline:2010kv,Cline:2012yx} also consider an exothermic model that lacks a threshold velocity entirely, which should lead to much cuspier profiles. On the other hand, Ref.~\cite{Pospelov:2007xh} considers a decaying metastable particle, which was found in Refs.~\cite{Ascasibar:2005rw,Abidin:2010ea,Vincent:2012an} to produce far too shallow an emission profile to fit the data. Future work that considers the uncertainty in the dark matter density profile could be valuable for comparing these different models.

Observations of other sources could help confirm this scenario. While the velocity dispersions in dwarf galaxies are likely too low to produce a signal from XDM, other galaxies and galaxy clusters offer a potential avenue for discovery. These remain out of reach of the sensitivity of INTEGRAL, but future experiments such as e-ASTROGAM \cite{e-ASTROGAM:2017pxr} and COSI \cite{Tomsick:2019wvo} may shed light on this scenario. Certain realisations of XDM may also be visible as inelastic dark matter in underground experiments. Although halo dark matter particles with $\sim$MeV mass splittings are out of kinematic reach, cosmic-ray boosted dark matter could be visible in heavy direct detection targets \cite{Bell:2021xff, Feng:2021hyz, Song:2021yar}. 

\acknowledgements
We thank Thomas Siegert for providing us with the INTEGRAL/SPI data used in this work, and Roland Crocker for helpful comments. This work is supported by the Natural Sciences and Engineering Research Council of Canada, the Arthur B. McDonald Canadian Astroparticle Physics Research Institute,  the Canada Foundation for Innovation and the Province of Ontario via an Early Researcher Award. Research at Perimeter Institute is supported by the Government of Canada through the Department of Innovation, Science, and Economic Development, and by the Province of Ontario.
 
\bibliography{main}

%%%%%%%%%%%%%%%%%%%%%%%%%%%%%%%%%%%%%%%%%%%%%%%%%%%%%%%%%%%%%%%%%%
%%%%%%%%%%%%%%%%%%%%%%%%%%%%%%%%%%%%%%%%%%%%%%%%%%%%%%%%%%%%%%%%%%

\end{document}